\documentclass{iaus}
\usepackage{graphicx}
\usepackage[dvips]{epsfig}

\title[The MDF of the Galactic Halo] 
{The Metallicity Distribution Function of the Halo of the Milky Way}

\author[Beers et al.]
 {Timothy C. Beers$^1$,
 Norbert Christlieb$^2$,
 John E. Norris$^3$,
 Michael S. Bessell$^3$,
 Ronald Wilhelm$^4$,
 Carlos Allende Prieto$^5$,
 Brian Yanny$^6$,
 Constance Rockosi$^7$,
 Heidi Jo Newberg$^8$,
 Silvia Rossi$^9$,
 Young Sun Lee$^1$}

\affiliation{$^1$Dept. of Physics \& Astronomy and\\ JINA: Joint Institute for
Nuclear Astrophysics,\\
Michigan State University (USA)\\
$^2$Hamburger Sternwarte, Univesit\"{a}t Hamburg, (Germany)\\
$^3$RSAA, Australian National University (Australia)\\
$^4$Texas Tech (USA)\\
$^5$McDonald Observatory, University of Texas (USA)\\
$^6$FermiLab (USA)\\
$^7$University of California, Santa Cruz (USA)\\
$^8$Rensselaer Polytechnic University (USA)\\
$^9$IAG, University of S\~{a}o Paolo (Brazil)}

\pubyear{2005}
\volume{228}  
\pagerange{1--7}
\date{?? and in revised form ??}
\jname{From Lithium to Uranium: Elemental Tracers of Early Cosmic Evolution}
\editors{V. Hill, P. Fran\c{c}ois \& F. Primas, eds.}
\begin{document}

\maketitle

\begin{abstract}
 
We report on the distribution of metallicities, [Fe/H], for very metal-poor
stars in the halo of the Galaxy. Although the primary information on the nature
of the Metallicity Distribution Function (MDF) is obtained from the two
major recent surveys for metal-poor stars, the HK survey of Beers and
collaborators, and the Hamburg/ESO Survey of Christlieb and collaborators, we
also discuss the MDF derived from the publicly available database of stellar
spectra and photometry contained in the third data release of the Sloan Digital
Sky Survey (SDSS DR-3). Even though the SDSS was not originally planned as a
stellar survey, significant numbers of stars have been observed to date -- DR-3
contains spectroscopy for over 70,000 stars, at least half of which are suitable
for abundance determinations. There are as many very metal-poor ([Fe/H] $<
-2.0$) stars in DR-3 as have been obtained from all previous survey efforts
combined. We also discuss prospects for significant expansion of the list of
metal-poor stars to be obtained from the recently funded extension of the SDSS,
which includes the project SEGUE: Sloan Extension for Galactic Understanding and
Exploration.


\end{abstract}

\firstsection 
\section{Introduction}

The distribution of stellar metallicities in the halo of the Galaxy has been
intensively studied for many years, especially since it was recognized in the
early 1980's that the halo contains field stars with [Fe/H] lower than that of the
lowest metallicity globular clusters. The central difficulty of this enterprise
is that low-metallicity stars, especially those with [Fe/H]~$< -2.0$, are
extremely rare in the solar neighborhood, comprising no more than $\sim 0.1$\% of the
stars within a few kpc of the Sun.  As a direct consequence,
knowledge of the form of the Metallicity Distribution Function (MDF), in
particular the shape of its low-metallicity tail, has long been limited by
small-number statistics. 

The difficulty of isolating a ``fair'' sample of halo stars based on in-situ
studies has left astronomers with little choice but to identify likely halo
stars based on their motions. Spectroscopic follow-up of stars selected from
proper-motion surveys has allowed the determination of at least a reasonably
accurate picture of the global shape of the halo MDF. The studies of Ryan \&
Norris (1991) and Carney et al. (1996) were able to demonstrate that the MDF of
halo stars peaks at a metallicity [Fe/H] $= -1.6$, including tails extending up
to the solar metallicity on the high side, and to metallicities at least down to
[Fe/H] = $-3.0$, or slightly below, on the low side. Both suggested MDFs are
consistent with one another, and with the predictions of the so-called ``Simple
Model'' (e.g. Hartwick 1976). The actual shape of the MDF at the lowest
metallicities, and its precise cutoff, is limited by the small numbers of stars
in these samples. Even taken together, these surveys only include some 250 very
metal-poor stars with [Fe/H] $\le -2.0$, and but a handful with [Fe/H] $\le
-3.0$.

Recent kinematically unbiased surveys for metal-poor stars have revealed the
presence of at least two hyper metal-poor (HMP) stars, with [Fe/H] $< -5.0$
(Christlieb et al. 2002; Frebel et al. 2005). The presence of these two stars
raise several additional questions concerning the nature of the MDF, including:

\bigskip

\begin{itemize}

\item What is the shape of the low-metallicity tail of the halo MDF, and in
particular, is it continuous ?

\item Does there exist a sharp cutoff of the halo MDF, and if so, at what
metallicity does it occur ?

\item Is the halo MDF constant throughout the halo, or does it vary with
distance ?

\end{itemize}

\bigskip

The answers to these, and other questions about the nature of the halo MDF, can
only be obtained from much larger samples of very metal-poor stars, ideally
chosen in-situ in the halo of the Galaxy, well outside the solar neighborhood.
Fortunately, due to the collective efforts of many of the astronomers present at
this conference, we have arrived at the point where the results of the
large-scale objective prism surveys (the HK survey of Beers and colleagues;
Beers et al. 1985, 1992, Beers 1999, and the HES of Christlieb and colleagues;
Christlieb 2003) can now begin to shed light on at least some of these
questions. Below we summarize the present picture of the low-metallicity tail of
the halo MDF as obtained by these two efforts. We then consider the information
provided by the large numbers of very metal-poor stars contained in the
most-recent public data release of the Sloan Digital Sky Survey (York et al.
2000), and its planned extension, SDSS-II, which includes the SEGUE program.   
  
\section{The HK Survey and the Hamburg/ESO Survey}

The HK survey (formerly known as the Preston-Shectman survey) was initiated over
25 years ago. During the course of the survey, a total of some 300
objective-prism plates covering 2800 deg$^2$ in the northern hemisphere and 4100
deg$^2$ in the southern hemisphere were obtained. The selection of candidate
metal-poor stars was accomplished based on visual inspection with a binocular
10~X microscope. Since the visual inspection was performed without the benefit of
the stellar colors (hence temperatures) it was expected that the HK-survey
candidates would carry a rather severe temperature-related bias. The potential
bias becomes less of a problem at the lowest metallicities, below [Fe/H] $=
-2.0$, where the CaII K lines of even quite cool stars are difficult to detect
at the resolution of the survey.

Medium-resolution (1-2 {\AA}) spectroscopic and broadband photometric follow-up
of candidate metal-poor stars from the HK survey has been underway for two
decades. For the majority of this period, the spectroscopic follow-up was
conducted with single-slit spectroscopy using 1.5m-2.5m telescopes at a variety
of observatories. Most recently, the use of the 6dF multiplexed spectrograph on
the UK Schmidt telescope made it possible to obtain more than one hundred such
spectra simultaneously, hence the pace of this effort was sped up considerably. 
Table 1 lists the total number of spectra of HK survey candidates (including
large numbers of field horizontal-branch and other A-type stars in addition to
the metal-poor candidates) obtained to date. This table also lists the numbers
of HK survey stars with available spectroscopy that have had optical broadband
photometry (primarily $UBV$) measured thus far. Near-IR $JHK$ from the
2MASS Point Source Catalog (Cutri et al. 2003) is also available for over 90\% 
of the HK-survey stars with available spectroscopy.

The HES objective-prism survey provides the opportunity to greatly increase the
number of very metal-poor stars identified by the HK survey. It reaches about
two magnitudes deeper than the HK survey ($B \sim 17.5$ vs. $B \sim 15.5$) and
also covers regions of the southern sky not sampled by the HK survey; a total of
about 8225 deg$^2$ of the sky above $|b| = 30^o$ is presently available from the
HES. Another 1275 deg$^2$ covered by the HES is now being scanned and analyzed
for candidates. The selection of metal-poor candidates from the HES database of
digital objective-prism spectra is performed using quantitative criteria including
automatic spectral classification (Christlieb, Wisotzki, \& Grasshoff 2002).
The strength of the CaII K line is determined using the $KP$ line index (Beers
et al. 1999), as measured directly from the objective-prism spectra. Due to the
broad wavelength range covered by the HES spectra it is also possible to
determine $B-V$ directly from the prism spectra, with an accuracy on the order
of $\sim 0.1$ mag. Stars that have CaII K lines weaker than expected for their
estimated $B-V$ colors and an approximate metallicity of [Fe/H] $< -2.5$ are
selected as low-metallicity candidates.  

Medium-resolution spectroscopic follow-up of candidate metal-poor stars from the
HES has been underway for the past five years, primarily with 2.5m-4m class
telescopes. The use of larger telescopes and substantially more efficient
spectrographs than were available for much of the HK survey follow-up results in
a much faster assembly of this information. The HES has also benefitted from
multiplex follow-up with the 6dF instrument. Table 1 summarizes the numbers of
HES targets (including stars other than metal-poor candidates) with available
medium-resolution spectroscopy obtained to date. Broadband optical photometry
(primarily $BVRI$) is only now being obtained for the metal-poor HES candidates
(Beers et al., in preparation); about 80\% of the stars with available
spectroscopy have $JHK$ from 2MASS.  All of the SDSS stars discussed below have available
$ugriz$ photometry. 

\begin{table}[tbp]
  \caption{\label{Tab:Observations} Observational Follow-Up of Surveys}
  \begin{center}
    \begin{tabular}{lrrrr}\hline

      Survey      & Spectra    &     Unique     &  $UBV$    &   $JHK$  \\\hline

      HK          & 14488      &    11212       &  4944     &   10438 \\
      HES         &  7465      &     6212       &   812     &    5078 \\
      SDSS-DR3    & 71396      &   $\sim 70000$ &  \dots    &   \dots \\\hline
    \end{tabular}
  \end{center}
\end{table}

\section{Stellar Data from the SDSS}

Many astronomers may not be aware that the publicly available medium-resolution
(2.5 {\AA}) spectroscopy and $ugriz$ photometry from the SDSS includes
substantial numbers of stars in the Milky Way. At present, through the DR3
release, these total over 70,000 stars. Analysis pipelines have been developed
(and are presently being refined), and value added catalogs that summarize best
estimates of atmospheric parameters (T$_{\rm eff}$, log g, [Fe/H]), as well as
radial velocities, distance estimates, and proper motions, are now being
assembled (Beers et al. 2004). By the time the final data from the main SDSS are
released (presently anticipated to be June 2006), the number of stars with
similar data should be well over 100,000. Most stars contained in the SDSS
database were not targeted specifically to be metal-poor halo objects, and
indeed they represent a complex assembly of objects selected for calibration and
reddening determinations, directed studies of various classes of stars (e.g. 
horizontal-branch stars, carbon stars, white dwarfs, late-type K and M dwarfs,
etc), as well as objects originally targeted as quasars that turned out to be
stars. 

In spite of its rather inhomogeneous assembly, the SDSS stellar database does
provide a useful means for sampling the tail of the halo MDF (though one of
course must be open to the issues of selection bias). 
 
\section{Metallicity Determinations for HK, HES, and SDSS stars}

The abundance calibration of Beers et al. (1999) is used to obtain estimates of
[Fe/H] for the HK and HES stars. This method relies on the variation of the CaII
$KP$ index with $(B-V)_0$ color as a function of [Fe/H]. This approach works well,
obtaining abundance errors on the order of $0.2-0.25$ dex over the color range
$0.3 \le (B-V)_0 \le 1.2$ and for stars with abundances $-4.0 \le {\rm [Fe/H]}
\le -0.5$. Near the red end of the color range, and near the high-metallicity
end as well, saturation of the CaII K lines becomes more of an issue; at these
extremes this method tends to underestimate the true metallicity of a given
star. At the very lowest metallicities, below [Fe/H]$ = -4.0$, the approach
could {\it over}-estimate the true metallicity, as a result of contamination of
the CaII K line by interstellar CaII, or by molecular carbon features in the
spectra of carbon-enhanced metal-poor stars. In fact, both of these situations
arose in the cases of the medium-resolution discovery spectra of the two known
HMP stars HE~0107-5240 and HE~1327-2326.

Not all of the HK/HES objects have available $(B-V)_0$ colors at present.  Thus,
we have developed techniques that can be used to estimate this color, based
on a calibration of the strength of the Balmer H$\delta$ line. 
Previous tests indicate that the colors estimated in this manner should be good
to on the order of about 0.03 mags over the range $0.3 \le (B-V)_0
\le 0.7$.  For cooler stars the rapidly declining strength of the H$\delta$
line with increasing $(B-V)_0$ renders the technique less useful. 
We are fortunate that many of the stars in the HK/HES surveys have 2MASS $JHK$
available.  This information is particularly useful for the redder stars that lack
$B-V$ photometry, as a newly designed calibration based on $KP$ and $(J-K)_0$
has been developed in order to provide estimates of [Fe/H] for such stars.
 
New methods have been developed for the estimation of metallicities (and other
atmospheric parameters) for stars in the SDSS spectroscopic database (Allende
Prieto et al. 2004, 2005), and are used in combination with the Beers et al.
(1999) and Wilhelm, Beers, \& Gray (1999) approaches in order to obtain
estimates of [Fe/H] that are accurate to on the order of 0.20 to
0.25 dex over the temperature range 4,000~K $\le {\rm T}_{\rm eff} \le $ 10000~K.

Carbon-Enhanced Metal-Poor (CEMP) stars, which are identified with large
frequency in all three surveys, present special challenges because their optical
colors can be confounded by the presence of molecular carbon features.  A new
calibration, based on the $KP$ and $GP$ indices and $(J-K)_0$, directly tied to stars
with a range of (externally known) [Fe/H] and carbon abundances, has been developed by
Rossi et al. (2005), and is used to estimate [Fe/H] and [C/Fe] for such stars.
   
\section{The ``As Observed'' MDF of the Halo}
 
Figure 1 (left panel) shows the distribution of [Fe/H] for some 6000 stars from
the HK survey that have measured $(B-V)_0$ or inferred colors that fall in the
range where our adopted calibrations can be applied. Note that the distinctly
bi-modal character of this distribution is simply the result of the imperfect
selection of candidate low-metallicity targets; the large number of stars with
[Fe/H $ > -1.5$ are the ``mistakes''. The 1200 stars with metallicities [Fe/H]
$\le -2.0$ are the sought-after objects.  

Figure 1 (right panel) shows a similar plot for the HES stars. Although the
total number of targets is only about half of the HK survey stars, the far
higher efficiency of the HES in the identification of very metal-poor stars has
greatly reduced the number of mistakes with [Fe/H] $> -1.5$. As a result, the
total number of HES stars with [Fe/H] $\le -2.0$ exceeds that of the HK
survey by some 300 stars. There are almost twice as many [Fe/H] $\le -3.0$ stars
found in the HES as have been identified in the HK survey.

\begin{figure}[tbp]
\begin{center} 
\epsfig{file=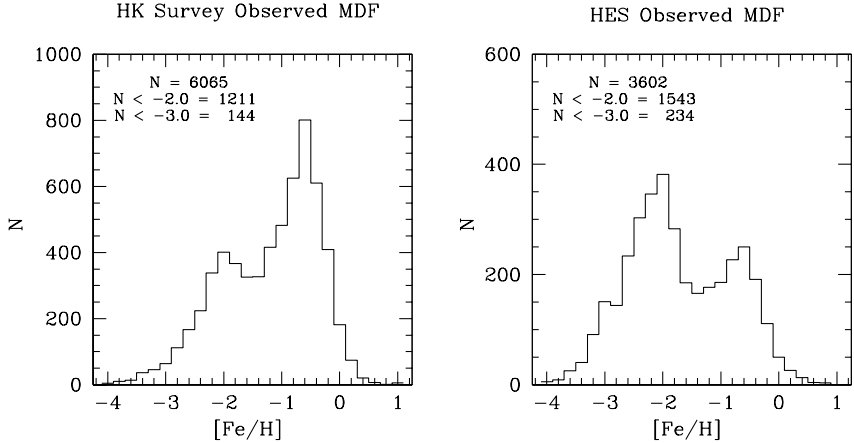, clip=, width=12cm}
\end{center}
\caption{The as-observed Metallicity Distribution Functions for the HK survey
(left panel) and the HES (right panel) candidate metal-poor stars. Bins are 0.2
dex in width. The numbers of stars with [Fe/H] $ < -2.0$ and [Fe/H] $< -3.0$ in
these surveys are listed individually. The selection efficiency of VMP stars in
the HES is clearly higher than that of the HK survey.}
\label{Fig:MDF}
\end{figure}

The combined samples of stars from the HK and HES includes
some 2700 stars with [Fe/H] $\le -2.0$ and almost 400 stars with [Fe/H]
$< -3.0$.  Note that we have not yet eliminated the HK-survey
stars that were rediscovered by the HES. Hence, these numbers will be reduced
somewhat in the final tally. 

There are two issues that need to be resolved before we can use the results of
these two surveys to place strong constraints on the shape of the low
metallicity tail of the halo MDF.  Clearly, owing to the manner in which the
surveys were constructed (to find the lowest metallicity stars), a bias is
introduced as metallicity rises above [Fe/H] $= -2.5$.  Our plan to correct for
this is to compare the shapes of the MDFs for stars in the range $-2.5 \le {\rm
[Fe/H]} \le -2.0$ with those of the kinematically-based surveys discussed
earlier, and derive a ``bias correction factor'' that can by applied to the
numbers of stars in this interval to account for the fraction lost by the
selection of candidates.  A similar exercise was performed for the HK survey
stars by Beers et al. (1998), who showed that the HK survey results available at
that time appeared to become biased above [Fe/H] $\simeq -2.2$. 

The second issue involves performing a careful check on the abundance estimates
for the many CEMP stars that are found at low metallicity in both surveys.
Experience has shown that it is possible to underestimate the metallicity of the
cooler, or more carbon-enhanced stars, even when making use of the new
techniques of Rossi et al. (2005). We therefore have to evaluate such stars on a
spectrum-by-spectrum basis, and either remove them or correct their abundances. 

Figure 2 shows the low-metallicity tail of the combined HK and HES MDFs, which
(for now) provides us with our best representation of the shape of this
function. Because of worries concerning the derived [Fe/H] for the CEMP stars
noted above, we show two versions of the combined MDF. The left panel is the
complete set of stars. The right panel is the same distribution, but with the
stars with estimated [C/Fe] $> +1.0$ removed. Two features are evident in these
diagrams. (1) The low-metallicity cutoff appears at [Fe/H]$ = -4.0$, although
the small numbers of stars with [Fe/H] $< -3.5$ still make the actual location
of the cutoff somewhat uncertain. The other feature is the possibility of a
small ``bump'' in the MDF at around [Fe/H] $= -2.5$. In order to be certain
about the presence of this bump, one would ideally like to obtain higher quality
medium-resolution spectra (and photometry where needed) of the HK/HES stars with
metallicities [Fe/H] $< -2.0$, using a single telescope/detector combination and
consistent S/N targets. Now that we know which stars to point at, and with the
use of high-efficiency modern spectrographs on 4m-class telescopes, this should
indeed be feasible on a reasonably short time scale. Note as well that, due to
the increase in the numbers of low-metallicity stars discovered over the course
of the HK/HES surveys that have now been studied at high spectroscopic
resolution, we have at our disposal a much larger set of objects with which one
could obtain an improved calibration of the metallicity-estimation procedures
employed in the original survey effort.

\begin{figure}[tbp]
\begin{center} 
\epsfig{file=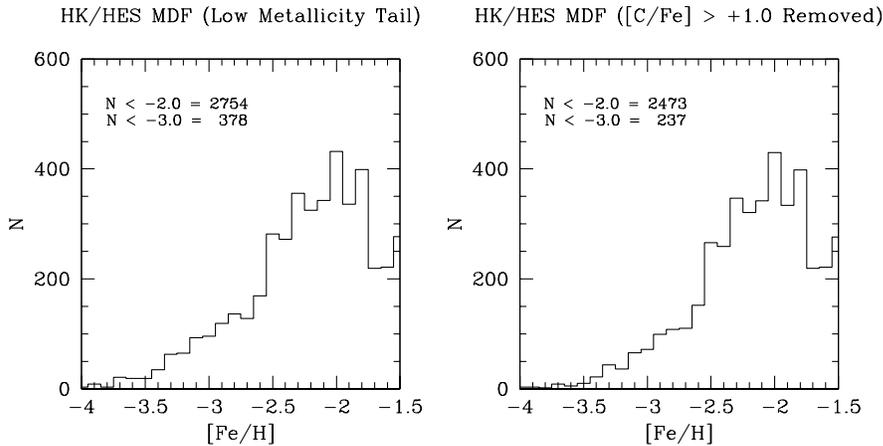, clip=, width=12 cm}
\end{center}
\caption{(left panel) The as-observed Metallicity Distribution Functions for the combined HK
survey HES candidate metal-poor stars, for stars with [Fe/H] $< -2.0$. (right
panel) The same distribution, but with the CEMP stars with [C/Fe] $> + 1.0$
removed.  Note that the bins are 0.1 dex in width. }
\label{Fig:MDFLOW}
\end{figure}

\section{A Look Ahead -- the Halo MDF from SDSS}

By the time of publication of this volume, we will be almost one-third of the
way through the next great expansion of the observational database of the
lowest-metallicity stars in the Galaxy. The survey effort, known as SEGUE, will
employ the same telescope (the ARC 2.5m telescope on Apache Point, New Mexico),
imager, spectrographs, and reduction pipeline as the original SDSS, to
dramatically increase our knowledge of the stellar populations of the Milky Way.
SEGUE, which begins in July 2005 and will end in July 2008, will obtain some
3500 additional square degrees of calibrated $ugriz$ photometry at lower
galactic latitude than the original SDSS, so as to better constrain the
important transition from the disk population(s) to the halo. Most importantly,
SEGUE will obtain medium-resolution spectroscopy for 250,000 stars in the
magnitude range $13.5 \le g \le 20.5$, in 200 directions covering the sky
available from Apache Point. SEGUE targets will be chosen to explore the Galaxy
at distances from 0.5 to 100 kpc from the Sun.

\begin{figure}[tbp]
\begin{center} 
\epsfig{file=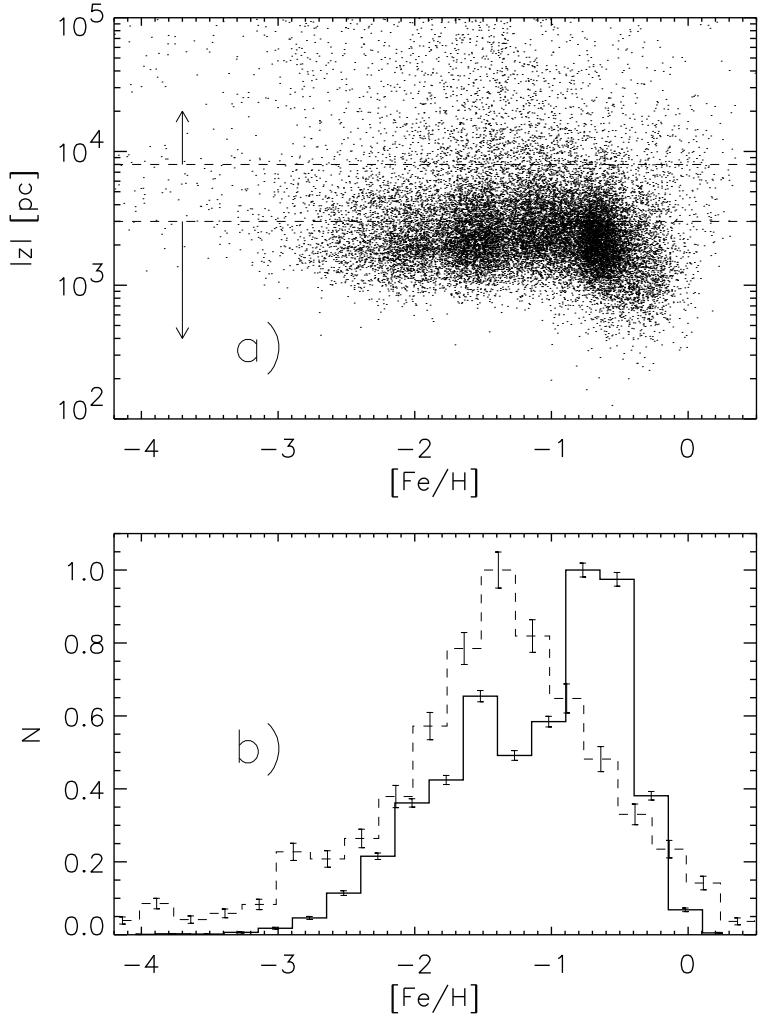, clip=, width=10 cm}
\end{center}
\caption{(a) MDF of SDSS-DR3 stars in the temperature range 5000~K $\le {\rm
T}_{\rm eff} \le 7000$~K, as a function of distance from the galactic plane.  The dashed lines
indicate the divisions in distance applied.  (b) MDFs of the stars from this
sample within 3 kpc from the plane (solid line) and those farther than 8 kpc
from the plan.  Bins are 0.2 dex in width. }
\label{Fig:MDFLOW}
\end{figure}

One of the primary SEGUE categories is obtained from photometric pre-selection
of likely very metal-poor (VMP) stars with [Fe/H] $< -2.0$. Tests carried out to
date indicate that it is reasonable to expect that SEGUE will yield a sample of
at least 25,000 VMP stars, a factor of ten times the present number of such
stars known from the summed HK and HES efforts. Although the majority of these
VMP stars will be too faint for easy high-resolution spectroscopic follow-up on
8m-10m class telescopes, there will be at least several thousand that {\it are}
sufficiently bright. Plans are presently being made to undertake HERES-like
``snapshot'' high-resolution spectroscopy (see Christlieb et al. 2004, Barklem
et al. 2005, and contributions from these authors in this volume) with the
Hobby-Eberly telescope, followed by higher S/N studies of the most interesting
objects that are revealed.

We can obtain a preview of the halo MDF that will be revealed by SEGUE from
inspection of the distribution of metallicities for stars that have been
observed (in a highly heterogeneous manner) during the course of the original
SDSS. Allende Prieto et al. (2005) have produced an ``as observed'' MDF for the
subset of SDSS stars in DR-3 with derived temperatures between 5000~K and
7000~K, where the abundance determination techniques developed for SEGUE are
expected to work best. Figure 3(a) shows the distribution of height above the
galactic plane as a function of estimated [Fe/H] for over 20,000 stars that fall
in this temperature interval. Figure 3(b) shows the MDFs for two sub-samples of
these stars, those with a distance from the galactic plane $|z| < 3$ kpc (14714
stars; solid histogram) and those with $|z| > 8$ kpc (2655 stars; dashed
histogram). The nearby cut in $z$ distance comprises a mix of halo and
thick-disk stars, while the more distant sample should be essentially all halo
stars.

It is remarkable how well the as-observed MDF of the SDSS DR-3 halo sample
mimics our expectation of what a ``fair'' sample of the halo MDF should look
like, though we re-emphasize that, owing to the peculiar selection of targets,
this sample should not be construed as such. There may be a few stars in DR-3
with metallicities below [Fe/H] = $-4.0$, but these objects will have to be
confirmed with high-resolution spectroscopy before a firm claim can be made as
to their existence. 

Note that, as part of the SEGUE target selection, there will be a very large ($N
\sim 100,000$) number of G-type dwarfs covering the distance range from a few
kpc out to 10-12 kpc from the Sun, and which are chosen in a ``metallicity
blind'' fashion. This sample will provide the opportunity not only to construct
an in-situ sample of stars from which a fair MDF of the halo can be obtained,
but also will enable determination of the fraction of VMP stars in
the solar neighborhood contributed by the Metal-Weak Thick-Disk population (see,
e.g. Chiba \& Beers 2000; Beers et al. 2002).

\section{Final Thoughts}

Clearly, the study of the MDF of the halo has entered a new era.  The prodigious
samples of halo stars that now exist from the HK survey, the HES, the SDSS, and
those that will be identified in the soon to-be-executed SEGUE survey are
finally producing sufficiently large numbers of objects to address the questions
raised above, as well as to (no doubt) raise many new ones.  Now is the time to
develop plans for the high-resolution spectroscopic follow-up of the many thousands of
SEGUE stars that will soon become available.

\begin{acknowledgments}

T.C.B. and Y.L. acknowledge support from grants AST 00-98508, AST 00-98549, AST
04-06784, and PHY 02-16783, Physics Frontier Centers/JINA: Joint Institute for
Nuclear Astrophysics, awarded by the US National Science Foundation. N.C.
acknowledges financial support through a Henri Chretien International Research
Grant administered by the American Astronomical Society, and from the Deutsche
Forschungsgemeinschaft under grants Ch~214/3 and Re~353/44. J.E.N. and M.S.B.
acknowledge support from grant DP0342613 awarded by the Australian Research
Council. C.A.P. acknowledges support from NASA grants NAG5-13057 and NAG5-13147.
C.R. acknowledges financial support from NASA grant HST-HF-01143.01-A. H.J.N.
acknowledges funding from the Research Corporation and from AST 03-07571,
awarded by the US National Science Foundation. S.R. thanks the Brasilian
Institutes FAPESP and CNPq for financial support.

\end{acknowledgments}


\begin{thebibliography}{}

\bibitem[]{} Allende Prieto, C. et al. 2004, in \textit{
Origin and Evolution of the Elements, Carnegie Observatories Astrophysics
Series}, eds A. McWilliam \& M. Rauch (Cambridge Univ. Press: Cambridge),
(astro-ph/0304352)

\bibitem[]{} Allende Prieto, C., et al. 2005, \textit{AJ}, submitted

\bibitem[]{} Barklem, P.S., et al. 2005, \textit{A\&A}, in press (astro-ph/0505050)

\bibitem[]{} Beers, T.C. 1999, in \textit{Third Stromlo Symposium: The Galactic Halo}, eds. B. Gibson,
T. Axelrod, \& M. Putman (ASP: San Francisco), 165, p. 206

\bibitem[]{} Beers, T.C., Preston, G.W., \& Shectman, S.A. 1985, \textit{AJ} 90, 2085

\bibitem[]{} Beers, T.C., Preston, G.W., \& Shectman, S.A. 1992, \textit{AJ} 103, 1987

\bibitem[]{} Beers, T.C., et al. 1998, \textit{Space Science Reviews} 84, 139

\bibitem[]{} Beers, T.C., et al. 1999, \textit{AJ} 117, 981

\bibitem[]{} Beers, T.C. et al. 2002, \textit{AJ} 124, 931

\bibitem[]{} Beers, T.C. et al. 2004, \textit{BAAS} 205, 21.08

\bibitem[]{} Carney, B.W., et al. 1996, \textit{AJ} 112, 668 

\bibitem[]{} Chiba, M., \& Beers, T.C 2000, \textit{AJ} 119, 2843

\bibitem[]{} Christlieb, N. 2003, \textit{Reviews in Modern
Astronomy} 16, 191

\bibitem[]{} Christlieb, N. Wisotzki, L., \& Grasshoff, G. 2002, \textit{A\&A} 391, 397

\bibitem[]{} Christlieb, N. et al. 2002, \textit{Nature} 419, 904

\bibitem[]{} Christlieb, N., et al. 2004, \textit{A\&A} 428, 1027

\bibitem[]{} Cutri, R. M., et al. 2003, \textit{VizieR Online Data Catalog}, 2246

\bibitem[]{} Frebel, A., et al. 2005, \textit{Nature} 434, 871

\bibitem[]{} Hartwick, F.D.A. 1976, \textit{ApJ} 209, 418

\bibitem[]{} Rossi, S., et al. 2005, \textit{AJ}, submitted

\bibitem[]{} Ryan, S.G., \& Norris, J.E. 1991, \textit{AJ} 101, 1865

\bibitem[]{} Wilhelm, R., Beers, T.C., \& Gray, R.O. 1999, \textit{AJ} 117, 2308

\bibitem[]{} York, D.G. et al. 2000, \textit{AJ} 120, 1579

\end{thebibliography}
\end{document}